\documentclass[3p,times,twocolumn]{elsarticle}

\usepackage{ecrc}

\volume{00}

\firstpage{1}

\journalname{Nuclear Physics B Proceedings Supplement}

\runauth{}

\jid{nuphbp}

\jnltitlelogo{Nuclear Physics B Proceedings Supplement}

\usepackage{amssymb}

\usepackage[figuresright]{rotating}

\begin{document}

\begin{frontmatter}



\dochead{}

\title{Neutrino searches with the IceCube telescope.}


\author{Juan A. Aguilar for the IceCube Collaboration.}

\address{D\'epartement de physique nucl\'eaire et corpusculaire, Universit\'e de Gen\'eve, CH-1211 Gen\`eve, Switzerland}

\begin{abstract}

The IceCube Neutrino Observatory is an array of 5,160 photomultipliers (PMTs) deployed on 86 strings at 1.5-2.5 km depth within the ice at the South Pole. The main goal of the IceCube experiment is the detection of an astrophysical neutrino signal. In this contribution we present the results of the point source analysis on the data taken from April 2008 to May 2011, when three detector configurations were operated: the 40-string configuration (IC-40), the 59-string configuration (IC-59) and the 79-string configuration (IC-79). No significant excess indicative of point sources of neutrinos has been found, and we present upper limits for an $E^{-2}$ muon neutrino flux for a list of candidate sources.  For the first time these limits start to reach $10^{-12}$ TeV$^{-1}$ cm$^{-2}$ s$^{-1}$ in some parts of the sky.

\end{abstract}


\end{frontmatter}


\section{Introduction}
\label{sec:intro}

The origin of Cosmic Rays (CRs) is still one of the unresolved questions of modern physics. The detection of a point source of neutrinos could not only prove the hadronic acceleration models responsible for CR emission but also identify the sources of these CRs. The IceCube neutrino telescope is located at the South Pole aiming at the detection of astrophysical neutrinos originating in the same environment as CRs. An array of 5,160 photomultipliers (PMTs) has been deployed in the antarctic ice in order to detect the Cherenkov light emitted by the secondary muon produced in muon neutrino interaction in the vicinity of the detector. The energy and direction of the secondary muon is inferred by the amount of light collected by the PMTs and the geometrical properties of the Cherenkov emission. This direction is, on average, very close to the direction of the parent neutrino, specially for $E_{\nu} >$ 10~TeV, and therefore the IceCube detector can work as a neutrino telescope. 
In Sec.~\ref{sec:det} of these proceedings we describe the three data samples used in this analysis. The likelihood method used for the search of point neutrino sources is explained in Sec.~\ref{sec:method} and the results are given in Sec.~\ref{sec:results}. Conclusions are given in Sec.~\ref{sec:conclusions}.

\section{Detector and event selection.}
\label{sec:det}

One of the most important steps in neutrino analysis is to remove the large atmospheric muon background. A significant part of this reduction is achieved at the South Pole by rejecting poorly reconstructed up-going events from the northern sky and selecting high
energy down-going muons in the southern sky. However, these online filters are not enough to obtain an event selection suitable for neutrino analysis. Further processing is needed including the addition of more CPU consuming track reconstructions and algorithms. In this analysis we use three different geometries of the IceCube detector over the last three years 
and therefore the event selections for these three periods are different. 

\begin{figure}[!t]
 \vspace{5mm}
\centering
\includegraphics[width=3.in]{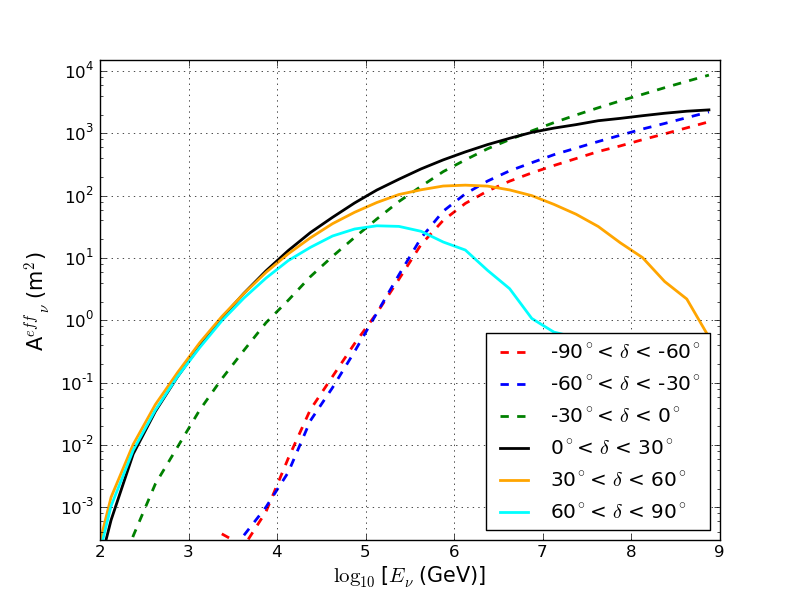}
\caption{Muon neutrino effective areas at final cut level for the IC-79 period for six declination bands (solid-angle-averaged) as a function of the neutrino energy.}
  \label{fig:aeff}
 \end{figure}

The event selection in IC-40 data was done by cutting on observables with discrimination power between signal (generally an E$^{-2}$ neutrino signal) and background (atmospheric neutrinos and muons). The details of this event selection can be found in Ref.~\cite{jonD}.  A multi-variate classification algorithm was used for the event selections in IC-59 and IC-79. A Boosted Decision Tree (BDT)~\cite{BDT} classifies events as signal-like or background-like based on a large number of detector observables. For the northern sky 12 and 17 variables were used in the BDT training for the IC-59 and IC-79 periods respectively.  The southern sky in the IC-59 dataset was, however, filtered using a cut on the energy estimator as a function of the zenith angle to separate the large amount of down-going atmospheric muons from a hypothetical neutrino signal with a harder spectrum. In the IC-79 dataset a combination of a BDT cut and a zenith dependent energy cut was applied. Figure~\ref{fig:aeff} shows the muon neutrino effective area of the IC-79 sample for different declination bands as a function of the neutrino energy. 

The different data rates for the three periods at different levels can be seen in Table~\ref{tab:rates}.

\begin{table}
\begin{center}
\begin{tabular}{l l l l l }
\hline
Conf. & Year & t$_{live}$ [days] & SMT [Hz] & $R_\nu$ \\
\hline
40 & 2008 & 376 & 1100 & 40/day\\
59 & 2009 & 348 & 1900 & 120/day\\
79 & 2010 & 316 & 2300 & 180/day\\
 \hline
\end{tabular}
\caption{Event rates for different IceCube configurations. First column is the number of strings in the configuration. The year of operation is shown in the second column. The live-time in days is given in the third column. The rate at trigger level (Simple Majority Trigger) is shown in the fourth column while the last column shows the number of up-going events (mainly atmospheric neutrinos) after the final cut.}
\label{tab:rates}
\end{center}
\end{table}

\section{Likelihood method for point source search}
\label{sec:method}

In this search an unbinned maximum likelihood ratio test~\cite{Braun} was used in order to calculate the significance of an excess of neutrinos above the background for a given direction in the sky. The signal ($S$) and background ($B$) probability density functions are modeled using simulation and scrambled real data, respectively. Since three geometries and event selections were combined, the density distribution of the $i^{th}$ event in the $j^{th}$ sample with energy $E_i$ and distance to the source $|x_i-x_s|$, is given by:

\begin{equation}
\label{eq:pdf}
P^j_i (|x_i-x_{s}|, E_i, \gamma, n^{j}_{s}) = \frac{n^{j}_{s}}{N^{j}}\mathcal{S}^{j}_{i} + \left(1- \frac{n^{j}_{s}}{N^{j}}\right)\mathcal{B}^{j}_{i},
\end{equation}

\noindent where $j$ represents the three different event selections, IC-40, IC-59 and IC-79. The total number of events in the $j^{th}$ data set, $N^{j}$, satisfies the condition $N_{tot} = N^{IC40}+N^{IC59}+N^{IC79}$, where $N_{tot}$ is the total number of events in the three samples. The signal is considered to follow the uniform hypothesis among data sets and therefore the spectral index meets the condition of $\gamma = \gamma_{j}$.  The fitted number of signal events $n^j_s$ in each sample are also fixed relative to each other, according to the signal hypothesis tested and the resulting fraction of total signal events expected in each sample. For the all-sky survey, the likelihood is evaluated in each direction in the sky in steps of 0.1$^{\circ} \times 0.1^{\circ}$ centered at the position of the source $x_s$. The significance of any excess found in the northern sky is corrected for trials by determining the fraction of scrambled data sets which contain an excess of equal or higher significance also in the northern part of the sky. Likewise, the most significant excess found in the southern sky is reported and the post-trial significance is calculated by comparing it with the fraction of scrambled data sets with an excess of equal or higher significance in that part of the sky.

\section{Results}
\label{sec:results}
\begin{figure*}[ht]
  \centering
  \includegraphics[width=5in,height=4in]{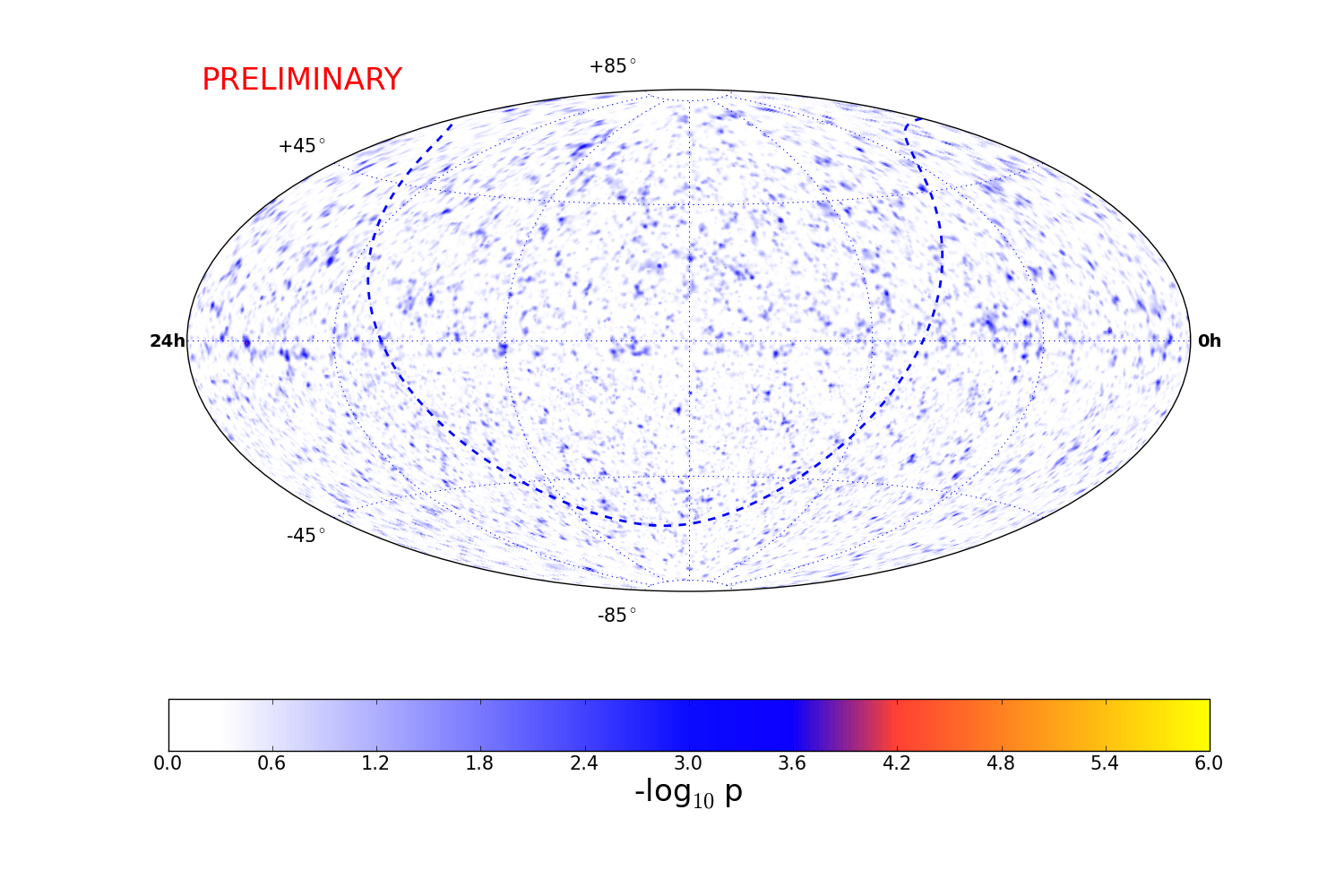}
  \caption{Significance skymap in equatorial coordinates (J2000) of the all-sky point source scan for the combined IC79+IC59+IC40 data sample. The dashed line indicates the galactic plane.}
  \label{skymap}
\end{figure*}

The results of the all-sky scan are shown in the pre-trial significance map in Figure~\ref{skymap}. The most significant deviation in the northern sky has a pre-trial $p$-value of 1.96~$\times$~10$^{-5}$ and is located at 34.25$^{\circ}$ r.a. and $2.75^{\circ}$ dec. Similarly, the most significant deviation in the southern sky has a pre-trial $p$-value of 8.97~$\times$~10$^{-5}$ and is located at 219.25$^{\circ}$ r.a. and $-38.75^{\circ}$ dec. The post-trial probabilities calculated as the fraction of scrambled sky maps with at least one spot with an equal or higher significance for each region of the sky correspond to 57\% and 98\% for the northern and the southern spots respectively and therefore both excesses are well compatible with the background hypothesis.

\begin{figure}[!t]
  \vspace{5mm}
  \centering
  \includegraphics[width=3.in]{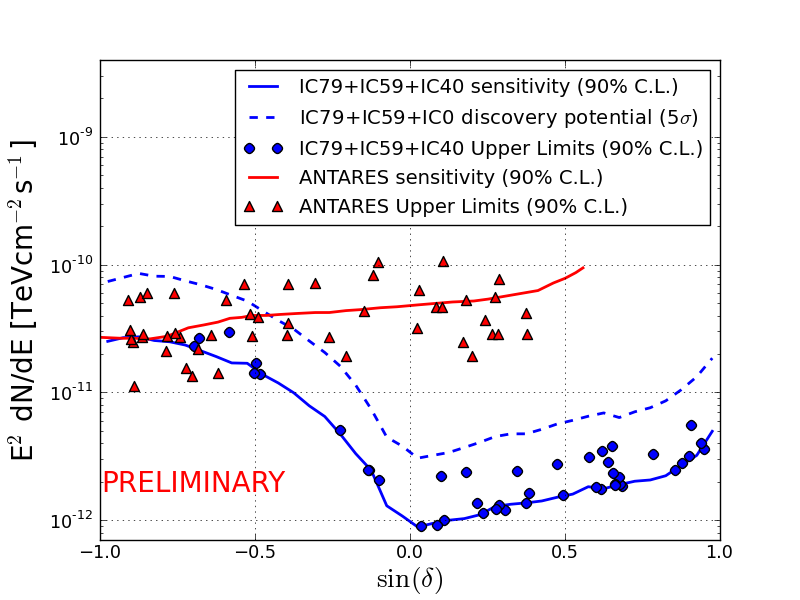}
  \caption{Muon neutrino flux 90\% C.L. upper limits and sensitivities for a $E^{-2}$ spectrum. Published limits of ANTARES~\cite{ANTARES} are shown.}
  \label{fig:ul}
 \end{figure}

Figure~\ref{fig:ul} shows the $E^{-2}$ muon neutrino flux upper limits calculated at 90\% C.L. based on the classical (frequentist) approach~\cite{Neyman} for each of the sources on a  candidate list.  Also shown is the median upper limit (sensitivity) for each declination for a live-time of 1,040 days and the discovery potential flux for a 5$\sigma$ confidence level. The 90\% C.L. muon neutrino upper limits and sensitivities from ANTARES~\cite{ANTARES} correspond to the analysis of 813 days of data.

\section{Conclusions}
\label{sec:conclusions}

We presented the results of the point source analysis using the data-sets corresponding to three different IceCube configurations. The combined data has a total live-time of 1,040 days. The search for neutrino point sources found no evidence of a neutrino signal. Both the north and the south hot spots have post-trial probabilities compatible with background fluctuations. The muon neutrino upper limits presented here are a factor $\sim 3.5$ better than the latest published upper limits by IceCube~\cite{jonD} and are the most competitive neutrino limits to date over the majority of the sky.




\nocite{*}
\bibliographystyle{elsarticle-num}



\end{document}